\documentclass[aps,floats,twocolumn,epsf,prm, superscriptaddress]{revtex4-1}

\usepackage{graphicx}
\usepackage{rotating}
\usepackage{amsmath}
\usepackage{amsfonts}
\usepackage{amssymb}
\usepackage{enumerate}
\usepackage{longtable}
\usepackage{dcolumn}
\usepackage{bm}
\usepackage{color}

\begin{document}

\newcommand{\be}{\begin{equation}}  
\newcommand{\ee}{\end{equation}}
\newcommand{\bn}{\begin{eqnarray}}
\newcommand{\en}{\end{eqnarray}}
\newcommand{\ii}{\'{\i}}
\newcommand{\ca}{\c c\~a}



\title{Electronic and valleytronic properties of crystalline boron-arsenide\\ tuned by strain and disorder}

\author{L. Craco}
\affiliation{Institute of Physics, Federal University of Mato Grosso, 78060-900, Cuiab\'a, MT, Brazil}

\author{S. S. Carara}
\affiliation{Institute of Physics, Federal University of Mato Grosso, 78060-900, Cuiab\'a, MT, Brazil}

\author{E. da Silva Barboza}
\affiliation{Institute of Physics, Federal University of Mato Grosso, 78060-900, Cuiab\'a, MT, Brazil}

\author{T. A. S. Pereira}
\affiliation{Institute of Physics, Federal University of Mato Grosso, 78060-900, Cuiab\'a, MT, Brazil}

\author{M. V. Milo\v{s}evi\'c}
\affiliation{Institute of Physics, Federal University of Mato Grosso, 78060-900, Cuiab\'a, MT, Brazil}
\affiliation{Department of Physics, University of Antwerp, Groenenborgerlaan 171, 2020 Antwerp, Belgium}

\begin{abstract}
\textit{Ab initio} density functional theory (DFT) and DFT plus coherent potential approximation (DFT+CPA) are employed to reveal, respectively, the effect of in-plane strain and 
site-diagonal disorder on the electronic structure of cubic boron arsenide (BAs). It is demonstrated that tensile strain and static diagonal disorder both reduce the semiconducting one-particle band gap of BAs, and a $V$-shaped $p$-band electronic state emerges - enabling advanced valleytronics based on strained 
and disordered semiconducting bulk crystals. At biaxial tensile strains close to $15\%$ the valence band lineshape relevant for optoelectronics is shown to coincide with one reported for GaAs at low energies. The role played by static disorder on the As sites is to promote $p$-type conductivity in the unstrained BAs bulk crystal, consistent with experimental observations. These findings illuminate the intricate and interdependent changes in crystal structure and lattice disorder on the electronic degrees of freedom of semiconductors and semimetals.
\end{abstract}

\maketitle

\section{Introduction}
\begin{figure}[b]
\includegraphics[width=0.8\linewidth]{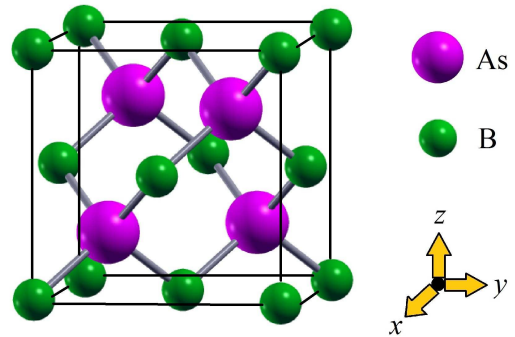}
\caption{Oblique view of the unit cell of the zinc-blende-type cubic BAs bulk crystal.} 
\label{fig1}
\end{figure}

Boron arsenide (BAs) crystallizes in the zinc-blende phase (depicted in 
Fig.~\ref{fig1}), which is isostructural with other III-V semiconductors such 
as GaAs, and is a $p$-type semiconductor as-grown~\cite{s10e11}. Compared to other III-V zinc-blende-type compounds, BAs had until recently not been considered suitable for applications due to difficulties with synthesizing high-quality single crystals. However, 
recent fabrication of highly pure samples of cubic BAs, and subsequent 
measurement of high room temperature thermal conductivity (in the 
range from 900 to 1300~W/mK~\cite{i5to7}), have placed BAs at the 
forefront of material research for future optoelectronic~\cite{bush,xmeng,islam} as well as 
photo-electrochemical~\cite{shi} applications. More precisely, 
within the context of electronic and optoelectronic materials, cubic BAs 
is expected to be useful for ultrahigh thermal management~\cite{fpan}, one of the prime challenges when operating electronic devices. Additionally, 
its low density, large resistivity, its isoelectronic configuration to that of Si, and feasible alloying with GaAs~\cite{zuger}, all open doors to versatile 
microelectronic applications of BAs~\cite{i15}. 

Experimental studies to date have indicated an inherent band gap between 1.46 and 
2.1~eV in cubic BAs bulk crystal~\cite{s12e13e14,shi,bsong,hlee}. Computational 
studies using density functional theory (DFT) revealed that the band structure 
of BAs is similar to other boron pnictides~\cite{houlong} as well as to 
GaAs~\cite{talwar}. In BAs the conduction band minimum occurs along 
the $\Gamma$-to-X direction and has $p$-band character, in contrast to AlAs, 
GaAs, or InAs, where the $s$-like conduction band minimum is found at the 
$\Gamma$-point. The difference may be attributed to the low energy position 
of the $p$ orbitals of boron ions~\cite{zuger}, as a result of strong 
covalence between B and As. Moreover, according to earlier DFT studies, 
the indirect band gap as well as the lowest direct 
band gap (at $\Gamma$) could be substantially different depending on the 
theory and approximations applied, with estimated values being between 0.7 and 
2.07~eV~\cite{scanlon,bsong}. Here, based on generalized gradient approximation 
(GGA) we estimate an indirect gap of 1.22~eV. This result agrees well with 
the values computed in earlier GGA calculations~\cite{anu,nwi},
which also slightly underestimate the band gap of the BAs bulk crystal.

For both fundamental and applied purposes it is worth noting here that 
the relationship between the energy of an electron and its momentum is 
determined by the band structure of the system, with the local minimum 
in the conduction band and local maximum in the valence band being referred 
to as {\it valley}~\cite{shai}, whose topology is governed by the crystal 
structure and chemical composition of the material. Thus, 
in addition to orbital, charge and spin, electrons in semiconductors and 
semimetals may also host valley degrees of freedom~\cite{zliu}. Similar 
to spin in spintronics~\cite{spintronics}, the valley degree of freedom opens the possibility to store 
and carry information, leading to an emerging device concept for electronic 
applications now known as valleytronics~\cite{valleytronics,ang}. In valleytronic materials~\cite{ang} electrons can populate low-energy 
valley states (or inequivalent local energy extrema)~\cite{shai}, 
providing an additional quantum index which introduces forefront paradigms 
to quantum information processing~\cite{sarma}. In addition to traditional 
semiconductors, such as AlAs~\cite{shko}, which have multiple valleys in the 
conduction bands located near the high symmetry points in the Brillouin zone, 
earlier works have reported tunable electronic 
structure and valley polarization of Dirac fermion systems~\cite{zliu}, 
such as graphene~\cite{gorbachev}, bismuth~\cite{feldman}, 
and Cd$_3$As$_2$~\cite{czhang}. One of the most documented tuning knobs is the lattice strain, widely used in the semiconductor electronics
industry to increase carrier mobilities~\cite{thomp,kyle}. In valleytronics, strain directly affects the valley-dependent energy dispersion~\cite{shai}, 
including manipulating the band gap size of conventional 
semiconductors~\cite{qin,oursupers,bush,xmeng} as well as of 
the spin-valley-coupled Dirac semimetals~\cite{zliu} with $V$-shaped band 
dispersion near the Fermi energy, $E_F$.  

Here, based on first principles band structure calculations we report the electronic structure reconstruction of an in-plane strained cubic BAs bulk crystal, exhibiting strain-tunable band gap narrowing, and emergence of a $V$-shaped $p$-band electronic dispersion suitable for advanced valleytronics. This study is timely, as straining BAs has become a viable technological approach. Namely, in their seminal work~\cite{ycui}, Cui \textit{et al.} recently demonstrated that self-assembled cubic BAs (also referred to as s-BAs) can be highly deformable,
to support uniaxial strains above 500$\%$ of its original size, similarly to homogeneous elastomers. In addition, the s-BAs can be compressed
or strained to random geometries without leading to a mechanical
breakdown, which recommends it for application as a thermal
interface material. Thus, a thermal interface based on s-BAs
with both high thermal conductivity and high flexibility is expected to find applications in flexible thermal cooling, soft robotics, among other emerging areas of flexible electronics~\cite{corzo}. 

Last but not least, it is of high relevance to above mentioned applications to understand the effect of the ever-present disorder in the experimentally synthesized BAs bulk crystals on the resulting electronic properties. For that purpose, 
we extend our first principles calculations to include the coherent potential approximation and examine the role of point defects on As sites in the electronic properties, with special attention paid to vacancies and antisites~\cite{qzeng} - being main actors as intrinsic lattice disorder in the BAs system~\cite{antisitedis}. Coupled to the observed tunability with strain, these results recommend BAs for further exploration towards advanced $p-$band electronics and valleytronics.

\section{Results and Discussion}

\subsection{Ab initio methodology and electronic structure of pristine cubic BAs}
Herein, we carried out DFT calculations to compute the band structure of 
pristine and strained BAs. The exchange and correlation part of the total 
energy was computed using GGA within the Perdew-Burke-Ernzerhof (PBE) 
functional~\cite{perdew} along with the DZP~\cite{wkohn} basis set. 
Non-relativistic pseudopotentials parameterized within the Troullier-Martins 
formalism were also used~\cite{klein}. As shown in our earlier studies, these 
combined approximations can accurately describe the electronic properties of 
pristine and strained graphene~\cite{our3759} as well as pressure-induced 
electronic structure reconstruction phosphorus allotropes~\cite{ourGP}. 
The unit cell of the cubic BAs consists of 8 atoms. For integrations 
over the Brillouin zone, a well-converged Monkhorst-Pack (MP) $k$-point 
mesh~\cite{monk} of $10 \times 10 \times 10$ (commensurating  with 
the lattice vectors) was used. We consider the cubic $F\bar 43m$ structure 
of BAs with two atoms in the primitive cell and with an experimentally 
reported lattice parameter of $a=4.777 \AA$~\cite{renata}. In order to find 
the equilibrium geometry of strained BAs, we minimize the energy with respect 
to the cubic lattice parameter $a$. Moreover, for strained BAs structural 
relaxation was performed using the conjugate-gradient method until the 
absolute value of the components of the Hellman-Feynman forces were converged 
to within 0.03~eV/$\AA$. The convergence with respect to the energy 
cut-offs were performed and the default energy cut-off value of 200~Ry was 
found to be sufficient. To establish a good accuracy between our results 
and computational costs, the tolerance in the matrix density was set to
$10^{-4}$~eV~\cite{lattice}. 

\begin{figure}[t]
\includegraphics[width=\linewidth]{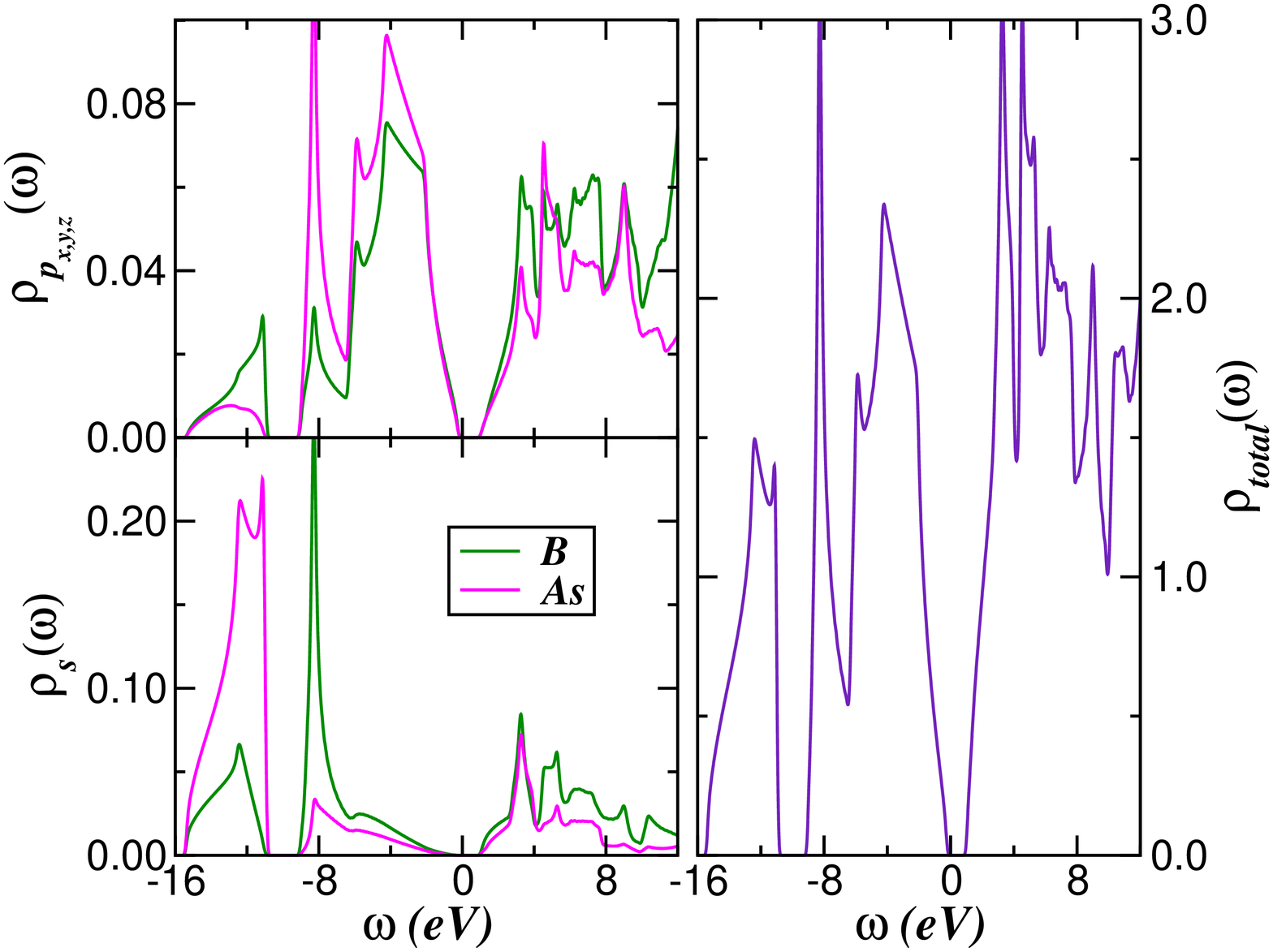}
\caption{ Partial (left) and total (right) density-of-states (DOS) of the cubic BAs bulk crystal. 
Notice the nearly similar $p$-band DOS near the Fermi energy $(E_F=\omega=0)$ 
due to large $p$-band hybridization.} 
\label{fig2}
\end{figure} 

In Fig.~\ref{fig2} we show the atom- and band-resolved density-of-states (DOS) 
of cubic BAs bulk crystal calculated using the GGA within the valence 
configurations $2s^22p^1$ and $4s^24p^3$ for B and As, respectively. As seen 
in the left upper panel, due to strong covalence the valence and conduction 
bands are nearly equally composed by both boron and arsenic $p$-type orbitals 
at low energies, with the valence band maximum almost reaching $E_F$. The 
conduction band is also mostly governed by threefold degenerate $p$-band 
electronic states, with the conduction band minimum being located at energies 
close to 1.2~eV above $E_F$. As visible in the lower left panel, valence band 
of cubic BAs has significant additional $s$-type contributions from both B 
and As at high binding energies. As mentioned above and consistent with 
earlier studies~\cite{scanlon,anu,john}, the minimum 1.22~eV band gap 
determined from the electronic band structure of unstrained BAs (not shown) 
is of indirect type. Finally, the $p$-type semiconducting nature intrinsic to 
cubic BAs bulk crystal is also manifested in total DOS (Fig.~\ref{fig2},
right panel), where small amount of hole doping would naturally introduce 
dominant $p$-band carriers in the electrical conductivity~\cite{john}.

\subsection{Tunability by strain}
It is now recognized that under external perturbations such as pressure or 
lattice strain the hopping elements of pnictides semiconductors and semimetals, including the one-particle band gap, can be renormalized in diverse and rather 
non-trivial ways~\cite{bush,houlong,zliu,tian}. We recall here, for example, 
the work by Liu {\it et al.}~\cite{zliu}, showing that the one-particle 
band gap of functionalized SbAs monolayers has a non-intuitive dependence on strain, where the band gap decreases gradually to zero and then 
increases with increasing elastic strain. Also interesting in this context is the study of
the BAs heterostructure by Bushick {\it et al.}~\cite{bush}, showing 
that isotropic biaxial in-plane strain and pressure decrease the band gap size. Moreover, based on atomistic calculations it has been 
predicted in Ref.~\onlinecite{bush} that $1\%$ biaxial tensile strain increases 
the in-plane electron and hole mobilities at 300 K by more than $60\%$ as 
compared to the unstrained BAs, due to reduced electron effective masses 
and hole interband scattering processes. Finally, the effect of biaxial strain on the electronic 
properties of hexagonal BAs single-layer ($h$-BAs) has been studied theoretically in Ref.~\onlinecite{kamara}, by
applying strains from 2 to $14~\%$, to reveal that this 
mono-layer material behaves as a direct band gap semiconductor for strains 
up to $8~\%$; the band gap of $h$-BAs changes from direct to indirect, and its size decreases with further increasing strain. Interestingly, 
this 2D-like system becomes metallic under 
biaxial strain of $14~\%$. Latter semiconductor-metal transition is caused 
by the lowering of the conduction band, reaching 
$E_F$ at the threshold strain of $\sim12~\%$. Below, we show that similar
electronic structure evolution is characteristic also of strained BAs bulk crystals. 

\begin{figure}[t]
\includegraphics[width=\linewidth]{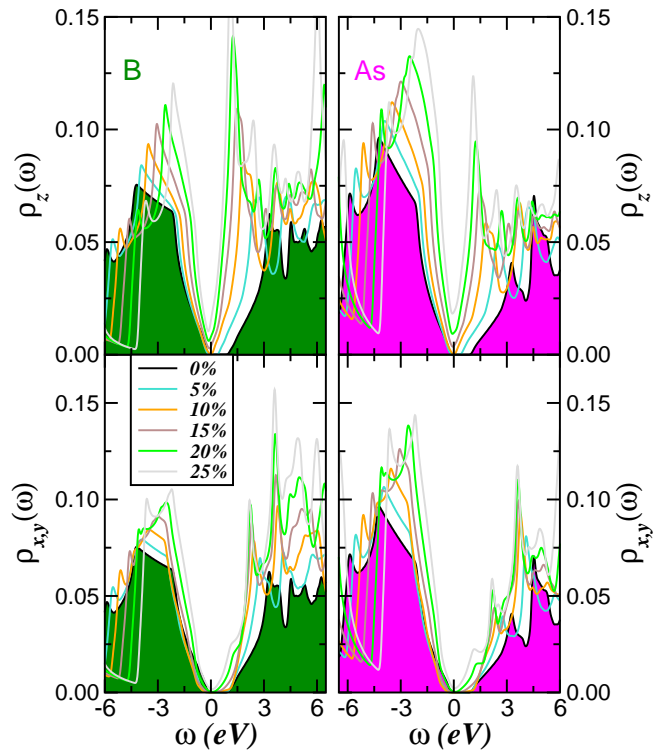}
\caption{Effect of strain on the atom- (left panel B, right As) and orbital-resolved $p$-band DOS of the BAs bulk crystal. Notice the lowering of the conduction band states and the systematic reduction of the band gap with increasing strain.} 
\label{fig3}
\end{figure}

\begin{figure}[t]
\includegraphics[width=\linewidth]{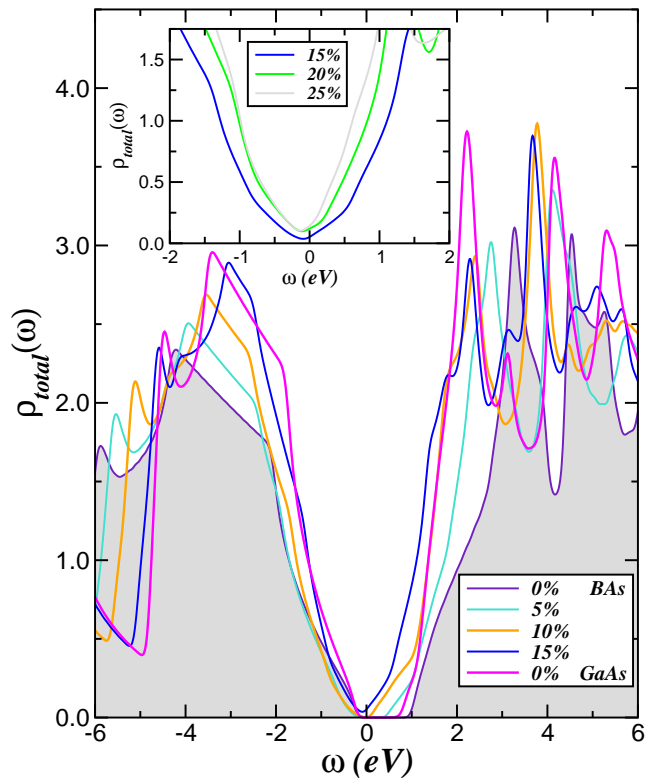}
\caption{Effect of biaxial lattice strain on the total DOS of a BAs bulk 
crystal. Notice the slightly larger band gap of BAs as compared to GaAs 
and the nearly similar valence (conduction) band low-in-energy lineshape 
between $15\%$ $(10\%)$ strained BAs and the unstrained GaAs semiconductor. 
Also noteworthy is the resulting nearly linear $V$-shaped spectra at large strains, shown in the inset.} 
\label{fig4}
\end{figure}

Fig.~\ref{fig3} shows the atom- and orbital-resolved $p$-band DOS of BAs 
as a function of the in-plane biaxial lattice strain. Several interesting and 
distinguishing features are seen in this figure, when compared to the 
unstrained spectral function. First, there is an overall downshift of the 
conduction band states toward lower energy values, while the valence band 
maximum is stabilized and almost pinned at energies very close to $E_F$. 
Second, the intensity of the lowest-energy van Hove-like peak in the 
$p_z$ DOS of B and As progressively grows with increasing strain. Third, 
and more surprisingly, a $V$-shaped like DOS characteristic of Dirac 
semimetals emerges near the semiconducting-semimetallic electronic transition, 
as seen in our results for the atom-resolved $p_z$ DOS at $10\%$ strain in 
Fig.~\ref{fig3}. Also remarkable is the lifting of the orbital degeneracy 
characteristic of cubic BAs bulk crystal, with the $p_z$ orbital showing 
distinct spectral features as compared to the twofold degenerate 
$p_{x,y}$ orbitals. As visible in Fig.~\ref{fig3}, the valence 
and conduction band spectra show appreciable one-particle band narrowing 
with increasing strain. Interesting as well, and consistent with earlier 
calculations for single-layer boron pnictides~\cite{houlong}, are our 
results for the total DOS in the main panel of Fig.~\ref{fig4}, showing 
how the occupied and unoccupied Dirac-like $V$-band lineshape progressively 
emerges in BAs with increasing in-plane biaxial strain.

To better visualize the systematic evolution of the nearly $V$-shaped
one-particle spectral function of strained BAs, in the inset of
Fig.~\ref{fig4} we display the total spectra for the in-plane biaxial
strain between 15 and 25$\%$. As seen, upon increasing strain a
consistent trend is observed in our reconstructed spectral functions, with
the $V$-like total DOS being less particle-hole asymmetric and closely
following a linear behavior in the energy range of $\pm 2.0$ eV. According
to this result, an all-electron semimetallic electronic state with
characteristic akin to  Dirac fermion systems~\cite{houlong} is
predicted to emerge in highly strained BAs bulk crystals.

\subsection{Assisted effect of disorder}
Owing to the importance of BAs bulk crystal for technological applications, 
we recall here that cubic BAs is known to incorporate high concentrations of 
crystal imperfections that can act as sources of $p$-type 
conductivity~\cite{john}. The latter has been attributed to the 
formation of native defects due to experimental grown conditions~\cite{gamage} 
as well as to unintentional acceptor impurities such as silicon and carbon, 
all potent to induce $p$-band conductivity. Interestingly, low temperature photoluminescence measurements of BAs bulk crystals revealed impurity-related recombination processes (including donor-acceptor pair recombination)~\cite{john}. It has also been found in Ref.~\onlinecite{john} that 
B impurities can be incorporated on the As sites, forming electrical 
antisite complexes. Finally, it has been reported theoretically that 
vacancies and antisites are rather typical point defects in BAs bulk 
crystals~\cite{qzeng}, both acting as intrinsic lattice disorder in the 
system~\cite{antisitedis}.

\begin{figure}[b]
\includegraphics[width=\linewidth]{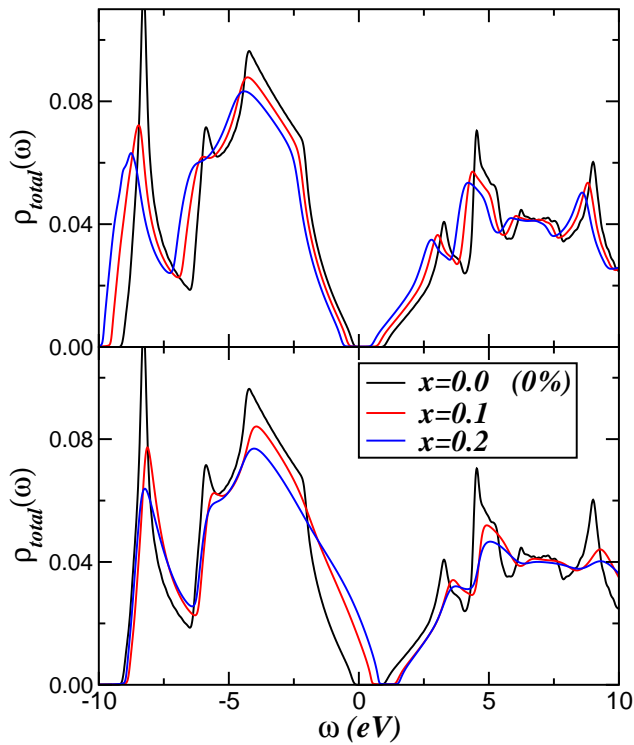}
\caption{Effect of site-diagonal disorder in the As total DOS of pristine BAs bulk crystal. Notice the changes in the electronic lineshape with increasing the concentration $x$ of local disorder. The lower and upper panels display, respectively, the effect of disorder due to B-antisites ($\delta=4.9$~eV) and As-vacancies ($\delta=-2.3$~eV).} 
\label{fig5}
\end{figure}

In what follows, we therefore focus on the effects of site-diagonal disorder 
in the As electronic states of a BAs bulk crystal. Here, we treat the 
disordered BAs within the DFT plus coherent potential approximation 
(DFT+CPA), which allows for an exact treatment of binary disorder in the 
limit of high lattice dimensions~\cite{DHM,cCPA}. Generally speaking, our 
scheme is the uncorrelated limit of that applied to the disordered Hubbard 
model~\cite{DHM}, where the effect of site-diagonal disorder is modeled by 
incorporating an Anderson-like disorder term 
$H_{dis} = \sum_{ia\sigma} v_{i} n_{ia\sigma}$~\cite{ourdisFeTe} in the bare 
one-electron Hamiltonian of BAs bulk crystal, which reads 
$H_{0}=\sum_{{\bf k}a\sigma} \epsilon_{a}({\bf k})c_{{\bf k}a\sigma}^{\dag}c_{{\bf k}a\sigma} 
+\sum_{ia\sigma} (\epsilon_{a}^{(0)} -\mu) n_{ia\sigma}$. Here, $a=x,y,z$ labels 
the diagonalized $p$-bands, $\epsilon_{a}({\bf k})$ is the one-electron band 
dispersion, which encodes details of the one-electron (GGA) band structure, 
$\mu$ is the chemical potential, and $\epsilon_{a}^{(0)}$ are on-site orbital 
energies of pristine and strained BAs, whose bare values are read off from the 
GGA spectral functions. In the spirit of Refs.~\onlinecite{DHM,volljarell} we 
restrict ourselves to a binary-alloy distribution for disorder, therefore 
the disorder potentials $v_{i}$ at the As sites are specified by the 
probability distribution $P(v_{i})=(1-x)\delta(v_{i}) + x\delta(v_{i}-\delta)$, 
meaning that upon incorporation of chemical disorder a fraction $x$ of As 
sites will have an additional local potential $\delta$ for an electron hopping 
onto that site. In other words, the As-$4p$ carriers of BAs bulk crystal 
experience different local environments in the course of their hopping, 
and the physical object which accounts for this effect is the 
CPA~\cite{DHM,anisV}. The detailed formulation of the disordered problem 
treated within CPA~\cite{DHM} has already been developed and used in the 
context of $p$-band semiconductors~\cite{ourGaAs} and 
semimetals~\cite{oursilicene}, so we do not repeat the equations here.

Fig.~\ref{fig5} shows the changes in the electronic structure of an unstrained
BAs bulk crystal, for different disorder parametrization. The $\delta$ values
used in this work are taken from the center of gravity of the B (4.9~eV)
and As (2.3~eV) total DOS, and in our CPA calculations they correspond to
antisite defects (i.e., B ions on As sites) as well to As vacancies,
whose deep-trap level~\cite{oba} is assumed to be the negative of As
center of gravity. The most salient feature to be seen in the panels
of Fig.~\ref{fig5} is that sizable electronic reconstruction is
predicted to emerge in unstrained BAs due to site-diagonal disorder
effects with increasing the concentration $x$ of disordered As
lattice sites. Particularly interesting is the emergence of $p$-type
conductivity~\cite{john} upon incorporation of B-antisite defects.
This is characterized by the appearance of low-energy electronic
states in the reconstructed DOS near $E_F$.

\begin{figure}[t]
\includegraphics[width=\linewidth]{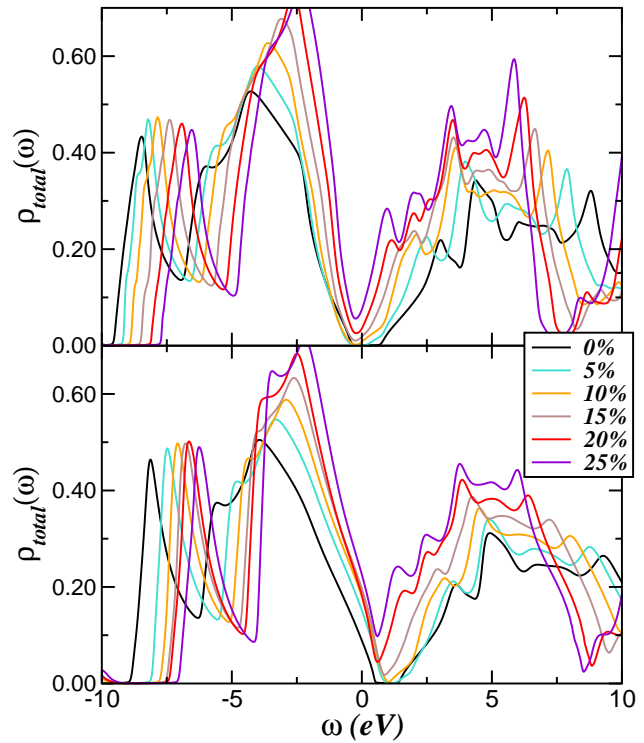}
\caption{Effect of site-diagonal disorder in the As total DOS, for 
$x=0.1$ and two different values of $\delta$ (two types of defects), 
for different values of biaxial strain from 5 to $25\%$. Notice the 
emergence of $p$- (lower panel) and $n$-type (upper panel) condutivity
due to B-antisites ($\delta=4.9$~eV) and As-vacancies ($\delta=-2.3$~eV), 
respectively. Noteworthy as well is the asymmetric $V$-shaped 
semimetal behavior in both panels.} 
\label{fig6}
\end{figure}

Even more interesting, however, are our results in Fig.~\ref{fig6}, showing
the changes in the electronic structure of disordered BAs bulk crystals
for the two values of $\delta$ mentioned above, for fixed disorder fraction $x=0.1$.
As seen, strong electronic reconstruction is predicted to emerge due
to the combined effect of site-diagonal disorder and lattice strain, 
inducing $p$- and $n$-band conductivity upon incorporation of
B-antisite defects (lower panel) and As-vacancies (upper panel). What lies at the
origin of these low-energy features? In a disordered system, incoherent
scattering between different carriers in orbital states leads to site-
and orbital-dependent shifts of the $p$-bands relative to each other,
and renormalized scattering rates~\cite{dobrohaule} due to sizable
$\delta$ cause appreciable spectral weight transfer over large energy 
scales, from high to low energies. This leads to a self-consistent 
modification of the spectral lineshape, as shown in Figs.~\ref{fig5}
and~\ref{fig6}. In accordance with
earlier studies~\cite{DHM,oursilicene,dobrohaule,mig}, broadening-induced
quantitative changes in the electronic structure reduce the band
gap and modify the low-energy electronic lineshape near $E_F$, both 
in normal (Fig.~\ref{fig5}) and highly strained (Fig.~\ref{fig6}) BAs
bulk crystal. As visible in Fig.~\ref{fig6}, the interplay between strain and
incoherent scattering arising from B-antisites and As-vacancies yields the
emergence of pseudogapped spectral functions in strained BAs with similarities
akin to disordered Dirac liquids: the incoherent semimetal found upon
application of strains above $10\%$ is thus predicted to be the counterpart
of the marginal Dirac liquid found in disordered silicene~\cite{oursilicene}.
More theoretical and experimental work on disordered semiconductors and
semimetals~\cite{svsy} are called for to corroborate our prediction of
strain-assisted $V$-shaped dirty Dirac-like liquids~\cite{durtyD}.

As a final remark, it is known that complementary electronics require both,
$p-$ and $n$-type semiconductors~\cite{meng}. Hence, according to our
results in Fig.~\ref{fig5}, disordered BAs at strains above $10\%$ are 
predicted to exhibit coexisting $p$- and $n$-type semiconducting 
characteristics, however, of ambipolar nature since the concentrations of electron and hole carriers would not be comparable. 
The implications of our results in Fig.~\ref{fig5} are rather remarkable in our opinion, since different transport polarities are predicted to exist 
in one single isostructural material system. According to our results, we 
propose that strained BAs with chemically engineered B-antisites and 
As-vacancies could feature intrinsic particle-hole asymmetric 
differential transconductance under positive and negative gate bias. 
Future studies should show whether this is in fact obtained in dirty BAs (or 
analogues) under sizable strain conditions. Once validated, our prediction 
may have far-reaching implications for integrated complementary 
electronics and data processing capability in a single 
device~\cite{meng}, thus significantly advancing the electronic circuit design.

\section{Conclusions}
In summary, using first-principles calculations augmented by coherent potential approximation for disorder, we have analyzed the strain- and disorder-induced electronic reconstruction and orbital differentiation of a BAs bulk crystal. We have revealed a variety of intricate one-particle effects that arise even with moderate strain values and disorder fractions. We find that both the in-plane strain (tensile or compressive) and disorder reduce the band-gap of BAs, and both promote the electronic state containing $V$-shaped Dirac-like valleys at the Fermi level. In absence of strain, the kind of defects at As sites (i.e. be it arsenic vacancies or boron substitutions) determines whether the conductivity will be of $p$ or $n$ type.

Since BAs bulk crystals are known to be stable at even extreme strains, and disorder in ever-present in grown crystals, our study provides a useful guide for future experiments. It also highlights the utility of first-principles and combined DFT+CPA calculations as fertile 
platforms to explore novel physical phenomena, including tuning the 
semiconducting band gap. This in turn opens prospects to further fundamental insights into localization-delocalization transition due to Anderson lattice disorder~\cite{anderson} in real $p$-band semiconductors and semimetals.

\acknowledgements  
This work was supported by Brazilian Agencies CNPq and CAPES, as well as by the Research Foundation-Flanders (FWO). L.C. thanks Byron Freelon and Stefano Leoni for discussion in earlier
stages of this work. E.S.B. acknowledges CAPES for individual financial support.


\begin{thebibliography}{28}
\bibitem{lattice} The calculated lattice parameters for strained 
BAs are: $a=b=5.0158~\AA,~ c=4.777~\AA$ ($5\%$ strain),
$a=b=5.255~\AA,~c=4.777~\AA$ ($10\%$ strain) and
$a=b=5.494~\AA,~c=4.777~\AA$ ($15\%$ strain).
\end{thebibliography}
\end{document}